\documentclass{PoS}                              % Documentstyle for PoS submission

\usepackage[latin1]{inputenc}                    % Codepage latin1
\usepackage{graphicx}                            % Graphics package
\usepackage{latexsym}                            % Load additional symbols
\usepackage{amsfonts}                            % Use AMS fonts,
\usepackage{amssymb}                             % symbols,
\usepackage{amsmath}                             % and definitions
\usepackage{dcolumn}                             % Align table columns
\usepackage{bm}                                  % Bold math

\graphicspath{{.}{Graphics/}{Figures/}}          % Path for graphics
\newcommand{\mps}{m_{\pi}}                       % m_\pi (pseudoscalar mass)
\title{Nucleon form factors on the lattice with light dynamical fermions}

\ShortTitle{Nucleon form factors}

\author{QCDSF/UKQCD Collaboration: M.~G{\"o}ckeler$^a$,
  Ph.~H{\"a}gler$^b$, R.~Horsley$^c$, Y.~Nakamura$^d$, M.~Ohtani$^a$,
  D.~Pleiter$^d$, P.~E.~L.~Rakow$^e$, A.~Sch{\"a}fer$^a$,
  G.~Schierholz$^{df}$, \speaker{W.~Schroers}$\,^d$\footnote{Current
    address: Institute of Physics, Academia Sinica, Nankang, Taipei
    115, Taiwan}, H.~St{\"u}ben$^g$, and J.~M.~Zanotti$^c$ \\
  \llap{$^a$}Institut f\"ur Theoretische Physik, Universit\"at
  Regensburg, 93040 Regensburg, Germany\\
  \llap{$^b$} Institut f{\"u}r Theoretische Physik T39,
  Physik-Department der TU M{\"u}nchen, James-Franck-Strasse,
  85747 Garching, Germany\\
  \llap{$^c$}School of Physics, University of Edinburgh,
  Edinburgh EH9 3JZ, UK\\
  \llap{$^d$}John von Neumann-Institut f\"ur Computing NIC /
  DESY, 15738  Zeuthen, Germany\\
  \llap{$^e$}Theoretical Physics Division, Department of Mathematical
  Sciences, University of Liverpool, Liverpool L69 3BX, UK\\
  \llap{$^f$}Deutsches Elektronen-Synchrotron DESY,
  22603 Hamburg, Germany\\
  \llap{$^g$}Konrad-Zuse-Zentrum f\"ur Informationstechnik Berlin,
  14195 Berlin, Germany}

\abstract{The electromagnetic form factors provide important insight
  into the internal structure of the nucleon and continue to be of
  major interest for experiment and phenomenology. For an intermediate
  range of momenta the form factors can be calculated on the lattice.
  However, the reliability of the results is limited by systematic
  errors mostly due to the required extrapolation to physical quark
  masses.  Chiral effective field theories predict a rather strong
  quark mass dependence in a range which was yet inaccessible for
  lattice simulations. We give an update on recent results from the
  QCDSF collaboration~\cite{Gockeler:2006uu,Gockeler:2006ui} using
  gauge configurations with dynamical $N_f=2$, non-perturbatively
  ${\cal O}(a)$-improved Wilson fermions at pion masses as low as 350
  MeV. \\ \\ Edinburgh 2007/22 \\ DESY 07-152 \\}

\FullConference{BARYONS 07\\
National Seoul University\\
Seoul, Republic of Korea}

\begin{document}

% --------------------------------------------------------------------------
%
%  Introduction
%
% --------------------------------------------------------------------------

\section{Introduction\label{sec:introduction}}
The proton and the neutron are the building blocks of atomic nuclei
and therefore the most important particles subject to the strong
interaction. Understanding their structure is therefore of strong
interest. For several decades they have been studied in detail and it
came as a great surprise at the beginning of this millennium that
experiments at Jefferson Laboratory deviated from the prevalent
theoretical understanding~\cite{Jones:1999rz}, see
also~\cite{Punjabi:2005wq} for recent reviews. The previous picture
was based on the perturbative behavior of the partons at
asymptotically high energies. A resolution of this mystery using
non-perturbative techniques is therefore in demand. Another puzzling
feature shows up when computing the size of the nucleon --- expressed
by its mean-squared radius, $\langle r^2\rangle$ --- within the
framework of chiral perturbation theory. In this scheme the size
diverges as the pion mass, $m_\pi$, decreases toward zero. Finding
the correct behavior again calls for the application of
model-independent non-perturbative techniques. Lattice QCD provides
such a description with the merit of being free from model assumptions
beyond QCD.

A further advantage of lattice QCD is the ability to vary the
parameters like $N_c$, $N_f$, and $m_q$. This allows for the validity
of specific model assumptions to be tested and is therefore of
importance also for the investigation of models of the strong
interaction, see e.g.~\cite{Goeke:2007fq}. Furthermore, in certain
cases experimental data is very hard to extract. Generalized parton
distributions~\cite{Mueller:1998fv} depend on several parameters and
their extraction from experiment relies on the applicability of QCD
factorization, see e.g.~\cite{Guidal:2007tc} for a discussion at this
conference. On the lattice, on the other hand, these observables can
be extracted without such difficulties and this has led to important
insights~\cite{Gockeler:2003jfa}.

On the downside, lattice simulations so far are limited to quark
masses above those in Nature. Decreasing the quark mass comes at the
expense of vastly increased demands in computer time. Despite
tremendous progress in recent years, simulating at the light quark
masses that Nature has chosen is still prohibitively expensive. To
meet this challenge, progress is required both in machine development
and in finding more efficient algorithms. To this day, even the most
advanced computations only reach quark masses corresponding to pion
masses of about 300 MeV and above.

To investigate hadron structure, three very different techniques are
being employed on the lattice today: (a) Ongoing simulations with
Wilson-type quarks down to smaller quark masses by exploiting more
efficient algorithms and increased computer
power~\cite{Gockeler:2006ns}. (b) Starting simulations with the
Ginsparg-Wilson formulation like domain-wall
fermions~\cite{Lin:2007gv} or overlap fermions~\cite{Cundy:2004xf}.
This approach has several advantages over Wilson fermions, for a
recent review consult~\cite{Bietenholz:2006ni}. It is, however, more
costly at the currently accessible quark masses. At heavy quark masses
these simulations are about 30 to 100 times more expensive. It is
furthermore computationally demanding since the entire parameter space
has to be explored again. (c) Using a hybrid-action approach with
different discretizations for the sea- and
valence-quarks~\cite{Negele:2004iu}. The latter approach constitutes a
compromise between low quark masses and performance at the expense of
conceptual uncertainties. The theory breaks unitarity at finite
lattice spacing which complicates the discussion of the continuum
limit. In practice, rooted staggered quarks are being used for the sea
quarks and it is not yet resolved if going to the continuum limit
commutes with taking the square root, see e.g.~\cite{Durr:2005ax} for
recent reviews.  Furthermore, matching sea- and valence quarks is
prescription-dependent and some prescriptions may give rise to
additional ${\cal O}(a^2)$ artifacts~\cite{Bar:2005tu}. Expressions
for finite-$a$ chiral extrapolations are known for several interesting
cases, but not all observables~\cite{Chen:2007ug}.

This paper presents recent results from the QCDSF collaboration on the
structure of the nucleon using two flavors of dynamical Wilson-Clover
fermions. This approach corresponds to the choice (a) above. The merit
is that it extends the existing data sets acquired over the past
decade using full QCD simulations. The parameter space is well
understood and surprises like unforeseen phase transitions are absent.
The questions addressed are the scaling behavior of the ratio of
$F_2(Q^2)/F_1(Q^2)$ at accessible values of the momentum transfer,
$Q^2\equiv -q^2=-t$, and the mean charge radii of the corresponding
form factors, $\langle r_1^2\rangle$ and $\langle r_2^2\rangle$,
together with the anomalous magnetic moment, $\kappa$. To set the
scale, we have set the Sommer parameter to a value of
$r_0=0.467$~fm. With this choice, the lattice spacings range from
$a=0.07\dots0.11$~fm and the pion masses cover a range of
$m_\pi=349\dots 1170$ MeV. We find that residual artifacts are small
compared to the statistical errors, so we analyze the data from
different lattice spacings together. The spatial volumes volumes vary
between $(1.4)^3\dots (2.6)^3$~fm$^3$. It is evident that the
parameter space is quite large and we address the question under which
circumstances chiral extrapolations can be attempted.

In this calculation we have restricted ourselves to the case of full
QCD, i.e., the sea- and valence quark masses are identical. The
extraction of the matrix elements has been discussed in detail in
previous publications~\cite{Gockeler:2003ay}. The renormalization has
been done non-perturbatively by requiring that the form factor
$F_1(Q^2=0)=1$ measures the electric charge of the proton.

% --------------------------------------------------------------------------
%
%  Numerical results
%
% --------------------------------------------------------------------------

\section{Numerical results\label{sec:numerical-results}}
\subsection{The ratio of spin-flip to spin-non-flip form
  factors\label{sec:ratio-spin-flip}}
\begin{figure}
  \centering
  \includegraphics[scale=0.3,angle=270,clip=true]{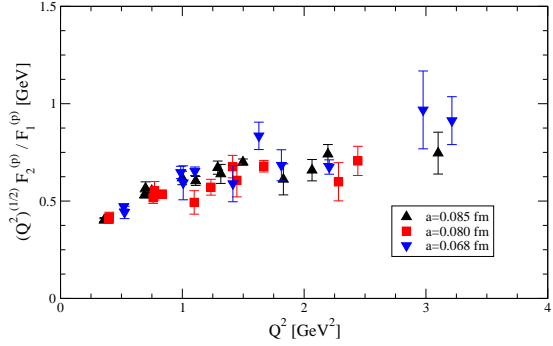}
  \caption{Momentum dependence of the form factor ratio. Results from
    three different lattice spacings at a pion mass $\mps\approx 600$
    MeV are shown.}
  \label{fig:scale-asymp}
\end{figure}

As has been mentioned in the introduction, the ratio of spin-flip to
spin-non-flip form factors, $F_2(Q^2)/F_1(Q^2)$, for sufficiently
large values of $Q^2$ was one of the key investigations in the recent
years. It was found that the experimental data is compatible with both
\begin{eqnarray}
  \label{eq:scaleasymp}
  \frac{\sqrt{Q^2}F_2(Q^2)}{F_1(Q^2)} &\to& \mbox{const}\,, \nonumber
  \\
  \frac{Q^2}{\log^2{Q^2}} \frac{F_2(Q^2)}{F_1(Q^2)} &\to&
  \mbox{const}\,.
\end{eqnarray}
We have tested this behavior using our lattice data at the accessible
values of $Q^2$. We find that the lattice data exhibits the same
behavior already at values of $Q^2>1\dots 2$ GeV$^2$. The results from
three different lattice spacings are displayed in
Fig.~\ref{fig:scale-asymp} for the first ratio in
Eq.~(\ref{eq:scaleasymp}). The result for the second ratio looks
similar is not displayed separately. The working points correspond to
a pion mass of $\mps\approx 600$ MeV. The behavior of the lattice data
is qualitatively consistent with the phenomenological
findings~\cite{Belitsky:2002kj}.

\subsection{Charge radii and the anomalous magnetic
  moment\label{sec:charge-radii-anom}}
Next, we discuss the charge radii, $\langle r_1^2\rangle$ and $\langle
r_2^2\rangle$, of $F_1$ and $F_2$ and the anomalous magnetic moment,
$\kappa$. In this work we restrict ourselves to the isovector case
$p-n=u-d$ since then the disconnected contributions cancel and a
comparison with experiment is free from any residual systematic errors
other than chiral, infinite volume and continuum extrapolations.

In order to parametrize the $Q^2$ dependence of the form factors, we
have adopted dipole and tripole-type fits~\cite{Gockeler:2006uu}. We
have also attempted to fit the form factors using a $p$-pole form with
a free parameter $p$
\begin{equation}
  \label{eq:p-pole}
  F_i(Q^2) = \frac{F_i(0)}{(1+Q^2/M^2)^p}\,.
\end{equation}
However, as has been shown in~\cite{Gockeler:2006uu}, we are unable to
measure the parameter $p$ to sufficient accuracy to distinguish
between dipole ($p=2$) and tripole ($p=3$) fits. In order to clearly
determine the optimal fitting form from lattice data alone a larger
range of $Q^2$ values is necessary. So we have to use additional
phenomenological input to specify our fitting formulae and perform
consistency checks.

In this work, we use the following fitting formulae:
\begin{eqnarray}
  \label{eq:fit-forms}
  F(Q^2) &=& \frac{F(0)}{(1+Q^2/M^2)^2}\,,\qquad
      \mbox{for $F_1^{u-d}(Q^2)$}\,, \nonumber \\
  F(Q^2) &=& \frac{F(0)}{(1+Q^2/M^2)^3}\,, \qquad
      \mbox{for $F_2^{u-d}(Q^2)$}\,.
\end{eqnarray}
In forthcoming publications~\cite{Schroers:Lat07,Schroers:2007ma} we
will report on different fit ans{\"a}tze. In this work, however, we
will just employ the dipole- and tripole-fits.

From these expressions, we can extract $\langle r_1^2\rangle$,
$\langle r_2^2\rangle$, and $\kappa$ by expanding
\begin{equation}
  \label{eq:charge-radii}
  F_i(Q^2) = F_i(0) \left( 1 - \frac{1}{6}\langle r_i^2\rangle Q^2 +
    {\cal O}(Q^4) \right) \,,\quad
  \langle r_i^2\rangle = \frac{6p}{M^2}\,, \quad
  F_2(0)=1+\kappa\,.
\end{equation}
To compare them with experiment, we need to perform a chiral
extrapolation. To this end, we discuss the formulae for the small
scale expansion (SSE) given in~\cite{Gockeler:2003ay}. Different
chiral expansions have been supplied in~\cite{Diehl:2006js}.

\begin{figure}
  \centering
  \includegraphics[scale=0.3,clip=true]{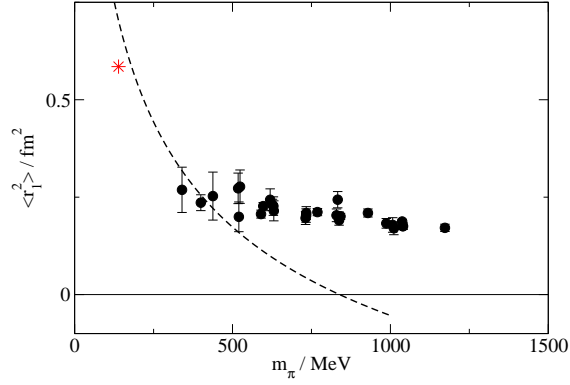}
  \caption{Physical value --- indicated by the star --- of $\langle
    r_1^2\rangle$ in the isovector case together with the lattice data
    and a chiral fit obtained from the SSE expression quoted
    in~\cite{Gockeler:2003ay}.}
  \label{fig:r1Q-vs-mps}
\end{figure}

When investigating the quantity $\langle r_1^2\rangle$, we fix the
appearing parameters to phenomenologically reasonable values. Note
that the radius diverges as $\mps$ approaches zero, implying that the
nucleon develops a larger and larger pion cloud as the quark mass goes
to zero. In the chiral limit the nucleon would be infinitely large. On
the other hand, the expression for $\langle r_1^2\rangle$ vanishes
already at a finite value of $\mps$ and becomes negative beyond. This
behavior is unphysical.  The resulting curve is displayed in
Fig.~\ref{fig:r1Q-vs-mps} together with the experimental data point
and our lattice results.
The lattice data describes a nucleon which is still smaller than the
experimental nucleon. Even at pion masses as low as $350$ MeV there is
no sign of a dramatic increase in size. The currently accessible pion
mass is still beyond the threshold of an expected sharp increase in
size.

\begin{figure}
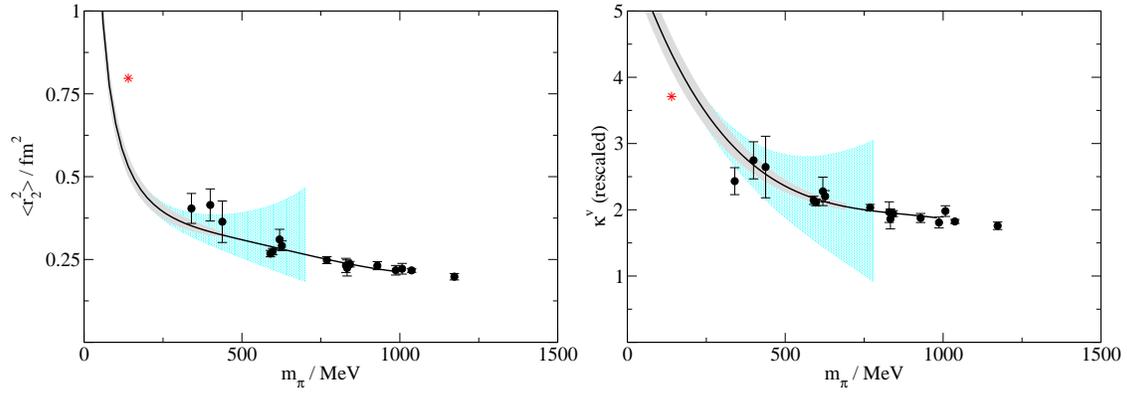

  \centering
  \includegraphics[scale=0.3,clip=true]{rv2Q.eps}
  \includegraphics[scale=0.3,clip=true]{kappav.eps}
  \caption{Combined fit to $\langle r_2^2\rangle$ (left) and $\kappa$
    (right). The fits as obtained from the SSE expressions
    from~\cite{Gockeler:2003ay} are shown together with the
    experimental results denoted by stars. The error bands are
    statistical (shaded) and systematical (dotted). $\kappa$ has been
    rescaled by the lattice nucleon mass as described
    in~\cite{Gockeler:2003ay}.}
  \label{fig:kr2-vs-mps}
\end{figure}

When turning to the radius $\langle r_2^2\rangle$ and the anomalous
magnetic moment $\kappa$, we find that the expansion of $r_2$
explicitly depends on $\kappa$. Hence, we can perform a joint fit of
both quantities and fix the parameters occurring in the expansion.
Altogether, $\kappa$ is taken to depend on three free parameters and
$r_2$ on the same three plus an additional constant. The results of
the combined fits are displayed in Fig.~\ref{fig:kr2-vs-mps}.
From these fits we conclude that the resulting values of $\langle
r_2^2\rangle$ and $\kappa$ are roughly compatible with the
experimental values at the physical pion mass. On the other hand, the
range of applicability of the chiral expansions does not seem to be as
large as one might have hoped. Further study of chiral expansions is
necessary to understand why these observables are described better
than $\langle r_1^2\rangle$.

% --------------------------------------------------------------------------
%
%  Summary
%
% --------------------------------------------------------------------------

\section{Summary\label{sec:summary}}
We have computed the behavior of the ratio $F_2(Q^2)/F_1(Q^2)$ of the
proton for different momentum combinations. We have also obtained the
charge radii $\langle r_1^2\rangle$ and $\langle r_2^2\rangle$, and
the anomalous magnetic moment, $\kappa$, of the nucleon for the
isovector combination $p-n=u-d$. Our calculation uses two flavors of
dynamical Wilson-Clover fermions and covers a large range of parameter
values down to pion masses of $350$ MeV. We find that the first
quantity is in qualitative agreement with the recent spin-transfer
experiments conducted at Jefferson Lab. Furthermore, the chiral
expansion together with the lattice data is consistent with
experimental values for the radius $\langle r_2^2\rangle$ and the
anomalous magnetic moment $\kappa$.

As of today lattice simulations are established as a reliable tool for
revealing the qualitative behavior of the structure of nuclear matter.
To perform a similar matching quantitatively from first principles
without additional model assumptions, however, will require more
progress both in lattice technology and in our understanding of chiral
expansions.  Nonetheless, we are confident that already by the end of
this decade quantitatively reliable predictions from first principles
will be available.

\acknowledgments The numerical calculations have been performed on the
Hitachi SR8000 at LRZ (Munich), the BlueGene/L and the Cray T3E at
EPCC (Edinburgh)~\cite{UKQCD}, the BlueGene/Ls at NIC/JFZ (J{\"u}lich)
and KEK (by the Kanazawa group as part of the DIK research program)
and on the APEmille and apeNEXT at NIC/DESY (Zeuthen). This work was
supported in part by the DFG under contract FOR 465 (Forschergruppe
Gitter-Hadronen-Ph{\"a}nomenologie and Emmy-Noether program) and by
the EU Integrated Infrastructure Initiative Hadron Physics (I3HP)
under contract number RII3-CT-2004-506078. W.S.~thanks Wolfgang
Bietenholz for valuable discussions.

% --------------------------------------------------------------------------
%
%  Bibliography
%
% --------------------------------------------------------------------------

% --------------------------------------------------------------------------
% 
%  End of document
% 
% --------------------------------------------------------------------------

\end{document}